\documentclass[5p,times]{elsarticle}

\usepackage{ecrc}

\journalname{}
\volume{99}
\firstpage{680}

\runauth{Y. H. Lam et al.}

\jid{}

\jnltitlelogo{}

\usepackage{amssymb}

\RequirePackage{booktabs}

\RequirePackage{tabularx}
\usepackage{longtable}
\newcommand{\mytilde}{{\raise.17ex\hbox{$\scriptstyle\mathtt{\sim}$}}}
\usepackage[pdftex]{hyperref}
\hypersetup
{
		bookmarks=true,%
		unicode=false,%
		pdftoolbar=true,
		pdfmenubar=true,
		pdffitwindow=false,
		pdfstartview=FitH,
		pdftitle={Yi Hua et Nadya proceeding},
		pdfauthor={LAM Yi Hua},
		pdfsubject={Nuclear Physics},
		pdfcreator={LAM Yi Hua},
		pdfproducer={LAM Yi Hua},
		pdfkeywords={Shell Model},		
        bookmarksnumbered,%
		colorlinks=true,%
		linkcolor=blue,%
		citecolor=blue,%
		filecolor=green,%
		urlcolor=blue,
		breaklinks, 
		plainpages=false%
}

\usepackage{CJKutf8}
\newcommand{\cntext}[1]{\begin{CJK*}{UTF8}{bkai}#1\end{CJK*}}

\usepackage[figuresright]{rotating}
\usepackage[dvipsnames]{xcolor}
\definecolor{mygray}{gray}{0.3}
\usepackage{amsmath}

\begin{document}

\begin{frontmatter}



\dochead{}

\title{Isobaric Multiplet Mass Equation for $A \le 71$ Revisited}
\tnotetext[1]{An electronic database of the isobaric multiplet mass equation coefficients is also available at \url{https://github.com/lamyihua/isobaric-multiplet-mass-equation}}
   

\author[1]{\textcolor{mygray}{Yi~Hua}~Lam (\cntext{藍乙華})}
\ead{lamyihua@impcas.ac.cn}
\ead[URL]{https://orcid.org/0000-0001-6646-0745}

\author[1]{\textcolor{mygray}{Bertram}~Blank}
\ead{blank@cenbg.in2p3.fr}
\author[1]{\textcolor{mygray}{Nadezda~A.}~Smirnova}
\ead[URL]{https://orcid.org/0000-0001-8944-7631}

\author[2]{\textcolor{mygray}{Jean Bernard}~Bueb}
\author[2]{\textcolor{mygray}{Maria Susai}~Antony}

\address[1]{CENBG (UMR 5797 --- Universit\'e Bordeaux 1 --- CNRS/IN2P3), \\ Chemin du Solarium, Le Haut Vigneau, BP 120, 33175 Gradignan Cedex, France}
\address[2]{IPHC, Universit\'e de Strasbourg, CNRS/UMR7178, 23 Rue du Loess, 67037 Strasbourg Cedex, France}

\begin{abstract}
Accurate mass determination of short-lived nuclides by Penning-trap spectrometers and progress in the spectroscopy of proton-rich nuclei have triggered renewed interest in the isobaric multiplet mass equation (IMME). The energy levels of the members of $T=1/2, 1, 3/2,$ and 2 multiplets and the coefficients of the IMME are tabulated for $A\le 71$. The new compilation is based on the most recent mass evaluation (AME2011) and it includes the experimental results on energies of the states evaluated up to end of 2011. Taking into account the error bars, a significant deviation from the quadratic form of the IMME for the $A=9, 35$ quartets and the $A=32$ quintet is observed.	
\end{abstract}


\end{frontmatter}



\section{Isobaric Multiplet Mass Equation}
\label{sec:IMME}

The isospin symmetry is an approximate symmetry of the nuclear Hamiltonian due to the presence of the electromagnetic interaction, the isospin non-conserving part of the strong force, and the difference in nucleonic masses. The largest source of the isospin symmetry breaking is the Coulomb interaction.	

In the isospin formalism, the Coulomb interaction has the form: 
\small
\begin{align}
\label{eq:vcoulomb1}
	V_{coul} &= \sum\limits_{i<j} \frac{Q_iQ_j}{| \vec{r}_i - \vec{r}_j|} \notag \\
             &= e^2 \sum\limits_{i<j} \left( \frac{1}{2} - t_z(i) \right) \left( \frac{1}{2} - t_z(j) \right) \frac{1}{| \vec{r}_i - \vec{r}_j|} \, ,
\end{align}
\normalsize
where $Q_i$ and $Q_j$ are the charge operators, $e$ is the electron charge, $t_z$ is the $z$-component of the isospin operator $\vec{t}$. Eq.(\ref{eq:vcoulomb1}) can be expanded as a sum of isoscalar, isovector, and isotensor operators ~\cite{Janecke1969, Frank_Jolie_VanIsacker2009},
\small
\begin{equation}
\label{eq:Vcoul}
	V_{coul} = \sum\limits_{q=0,1,2} V_{coul}^{(q)} \, ,
\end{equation}
\normalsize
namely,
\small
\begin{align}
\label{eq:Vcoul_comp}
	V_{coul}^{(0)} &= e^2 \sum\limits_{i<j} \left( \frac{1}{4} + \frac{1}{3} \vec{t}(i) \cdot \vec{t}(j) \right) \frac{1}{| \vec{r}_i - \vec{r}_j|} \, ,\notag \\ 
	V_{coul}^{(1)} &= - \frac{e^2}{2} \sum\limits_{i<j} \left( t_z(i) + t_z(j) \right) \frac{1}{| \vec{r}_i - \vec{r}_j|} \, ,\notag \\ 
	\textnormal{and } \notag \\
    V_{coul}^{(2)} &=  e^2 \sum\limits_{i<j} \left( t_z(i)t_z(j) - \frac{1}{3} \vec{t}(i) \cdot \vec{t}(j) \right) \frac{1}{| \vec{r}_i - \vec{r}_j|} \, .
\end{align}
\normalsize
All the operators are $0$-components of rank-$q$ isotensors, since the electric charge is conserved.

The nuclear Hamiltonian $H_0$ (Coulomb excluded) is isospin invariant, and has an SU(2) symmetry in isospace. Its eigenstates are characterised by the total isospin quantum number $T$, and for each $T$ they are $(2T+1)$-fold degenerate ($T_z = -T, -T+1, \ldots, T-1, T$), forming an isobaric multiplet of states. These states are denoted as $\left| \alpha ,T , T_z \right\rangle$, where $\alpha$ refers to other relevant quantum numbers ($J$, $\pi, A,\ldots$).

The effect of the Coulomb interaction can be treated within perturbation theory. In the lowest-order approximation, the isospin symmetry turns out to be broken in a dynamical way ~\cite{Frank_Jolie_VanIsacker2009}. The isoscalar term $V_{coul}^{(0)}$ is invariant with respect to the isospin SU(2) group, whereas the isovector term $V_{coul}^{(1)}$ and the isotensor term $V_{coul}^{(2)}$ are invariant with respect to the isospin group SO(2), a subgroup of SU(2). $V_{coul}^{(1)}$ and $V_{coul}^{(2)}$ contain the operators $T_z$ and $T_z^2$, respectively. 

In the lowest order of perturbation theory, the energy shift of a given member of an isobaric multiplet due to the Coulomb interaction is expressed by the expectation value of the Coulomb interaction in this state, namely,
\small
\begin{equation}
\label{eq:Ecd1}
	E_{coul}(\alpha, T, T_z) = \left\langle \alpha, T, T_z \right| V_{coul} \left| \alpha ,T , T_z \right\rangle  \, .
\end{equation}
\normalsize
Applying the Wigner-Eckart theorem ~\cite{Edmonds1957}, we can factor out the $T_z$ dependence, to obtain the following expression: 
\small
\begin{align}
\label{eq:Ecd2}
	& E_{coul}(\alpha, T, T_z) \notag \\
	&= \left\langle \alpha, T, T_z \right| \sum\limits_{q=0,1,2} V_{coul}^{(q)} \left| \alpha ,T , T_z \right\rangle \notag \\
	&=	\sum\limits_{q=0,1,2} (-1)^{T-T_z} \begin{pmatrix} T & q & T \\ -T_z & 0 & T_z \\ \end{pmatrix} \left\langle \alpha, T \right\| V_{coul}^{(q)} \left\| \alpha ,T \right\rangle \notag \\
	&= E_{coul}^{(0)}(\alpha, T) + E_{coul}^{(1)}(\alpha, T) T_z + E_{coul}^{(2)}(\alpha, T) (3T_z^2 - T(T+1)) \, ,
\end{align}	
\normalsize
where 
\small
\begin{align}
\label{eq:Ecoul2}
	E_{coul}^{(0)}(\alpha, T) &= \frac{1}{\sqrt{2T+1}} \left\langle \alpha, T \right\| V_{coul}^{(0)} \left\| \alpha ,T \right\rangle \, , \notag \\
	E_{coul}^{(1)}(\alpha, T) &= \frac{1}{\sqrt{T(2T+1)(T+1)}} \left\langle \alpha, T \right\| V_{coul}^{(1)} \left\| \alpha ,T \right\rangle \, , \notag \\
	E_{coul}^{(2)}(\alpha, T) &= \frac{1}{\sqrt{T(2T+3)(2T+1)(T+1)(2T-1)}} \notag \\ 
    & \left\langle \alpha, T \right\| V_{coul}^{(2)} \left\| \alpha ,T \right\rangle \, .
\end{align}
\normalsize
The matrix elements with two bars denote reduced matrix elements in isospin space.

Thus, in lowest-order perturbation theory, the diagonal matrix elements of the Coulomb interaction \\$\left\langle \alpha, T, T_z \right| V_{coul} \left| \alpha ,T , T_z \right\rangle$ are given by a quadratic form of $T_z$, c.f. Eq.~(\ref{eq:Ecd2}), while the off-diagonal isospin mixing matrix elements of 
$\left\langle \alpha, T, T_z \right| V_{coul} \left| \alpha ,T' , T_z \right\rangle$ are neglected. In this case, $V_{coul}$ is assumed not to mix states $\left| \alpha ,T , T_z \right\rangle$ having different values of $T$ ($T=T_z, T_z +1, \ldots $), and the isospin $T$ is still a good quantum number. However, the $(2T+1)$-fold degeneracy is now removed. The isobaric multiplet is thus split into $(2T+1)$ components.

We make use of Eq.~(\ref{eq:Ecd2}) to obtain the mass excess of an isospin-$T$ multiplet member in a specific state defined by $\alpha $: 
\small
\begin{align}
\label{eq:IMME2}
	M(\alpha, T, T_z) &= \frac{1}{2} (M_n + M_H)A + (M_n - M_H) T_z \notag \\
	&+ \left\langle \alpha, T , T_z \right|H_0 \left| \alpha ,T , T_z \right\rangle + E_{coul}(\alpha, T, T_z) 
\end{align}
\normalsize
where $M_n$ and $M_H$ are the neutron and the hydrogen mass excesses, respectively ({\it mass excess} is the difference between the experimental nuclear mass and $A$ atomic mass units, expressed in energy units here). $H_0$ represents the isospin-invariant nuclear Hamiltonian having charge independent interactions only. Therefore, we can rewrite Eq.~(\ref{eq:IMME2}) as 
\small
\begin{equation}
\label{eq:IMME}
	M(\alpha, T, T_z) =	a(\alpha, T) + b(\alpha, T) T_z + c(\alpha, T) T_z^2 \, ,
\end{equation}	
\normalsize
which is the {\it isobaric multiplet mass equation}, IMME ~\cite{Wigner58, WeinbergTreiman1959}, where
\small
\begin{align}
\label{eq:IMME_abc}
	a(\alpha, T) &= \frac{1}{2} (M_n + M_H)A 
				 + \left\langle \alpha, T , T_z \right|H_0 \left| \alpha ,T , T_z \right\rangle \notag \\
				 &+E_{coul}^{(0)}(\alpha, T) - T(T+1) E_{coul}^{(2)}(\alpha, T) \, , \notag \\
	b(\alpha, T) &= \Delta_{nH} - E_{coul}^{(1)}(\alpha, T) \, , \notag \\
\textnormal{ and }	\notag \\
    c(\alpha, T) &= 3 E_{coul}^{(2)}(\alpha, T) \, .
\end{align}
\normalsize
The neutron-hydrogen mass (or mass excess) difference is $\Delta_{nH}=M_n -M_H=782.34664$ keV. 

Obviously, charge-dependent forces of nuclear origin of a two-body type can be treated in the same manner, so the form of Eq.~(\ref{eq:IMME}) stays valid.

\begin{figure*}[ht!]
\centering
\rotatebox[]{0}{\includegraphics[scale=0.5, angle=0]{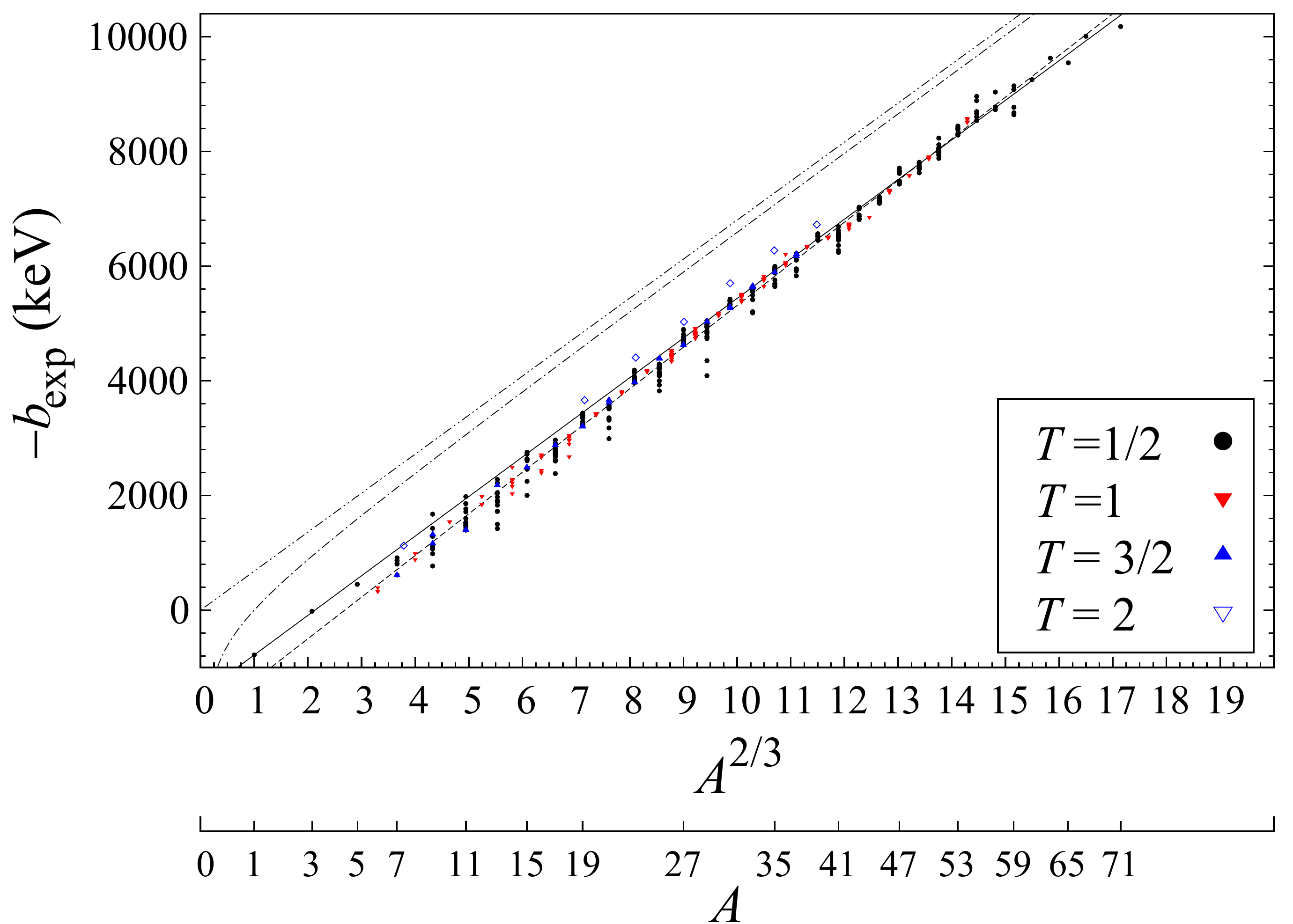}} 
\caption{\label{fig:All_b_coeff}
\footnotesize
The $b$ coefficients of the quadratic IMME (\protect\ref{eq:IMME}) as a function of $A^{2/3}$ for all $T=1/2,1,3/2$, and 2 multiplets.
		A weighted fit to $b$ coefficients, $b=-690.98(\pm89) A^{2/3} + 1473.02(\pm93)$ (keV) is displayed by the solid line. The dashed line shows the unweighted fit to $b$ coefficients, $b=-726.64 A^{2/3} + 1952.7$ (keV). The dashed-dotted line is $b=-\frac{3 e^2 (A-1)}{5 r_0 A^{1/3}}$. The double-dotted-dashed line is $b=-\frac{3 e^2}{5 r_0}A^{2/3}$.}
\end{figure*}

\section{Compilation of IMME}
\label{sec:IMME_Fitting_Procedure}

\begin{figure*}[ht!]
\centering
\rotatebox[]{0}{\includegraphics[scale=0.5, angle=0]{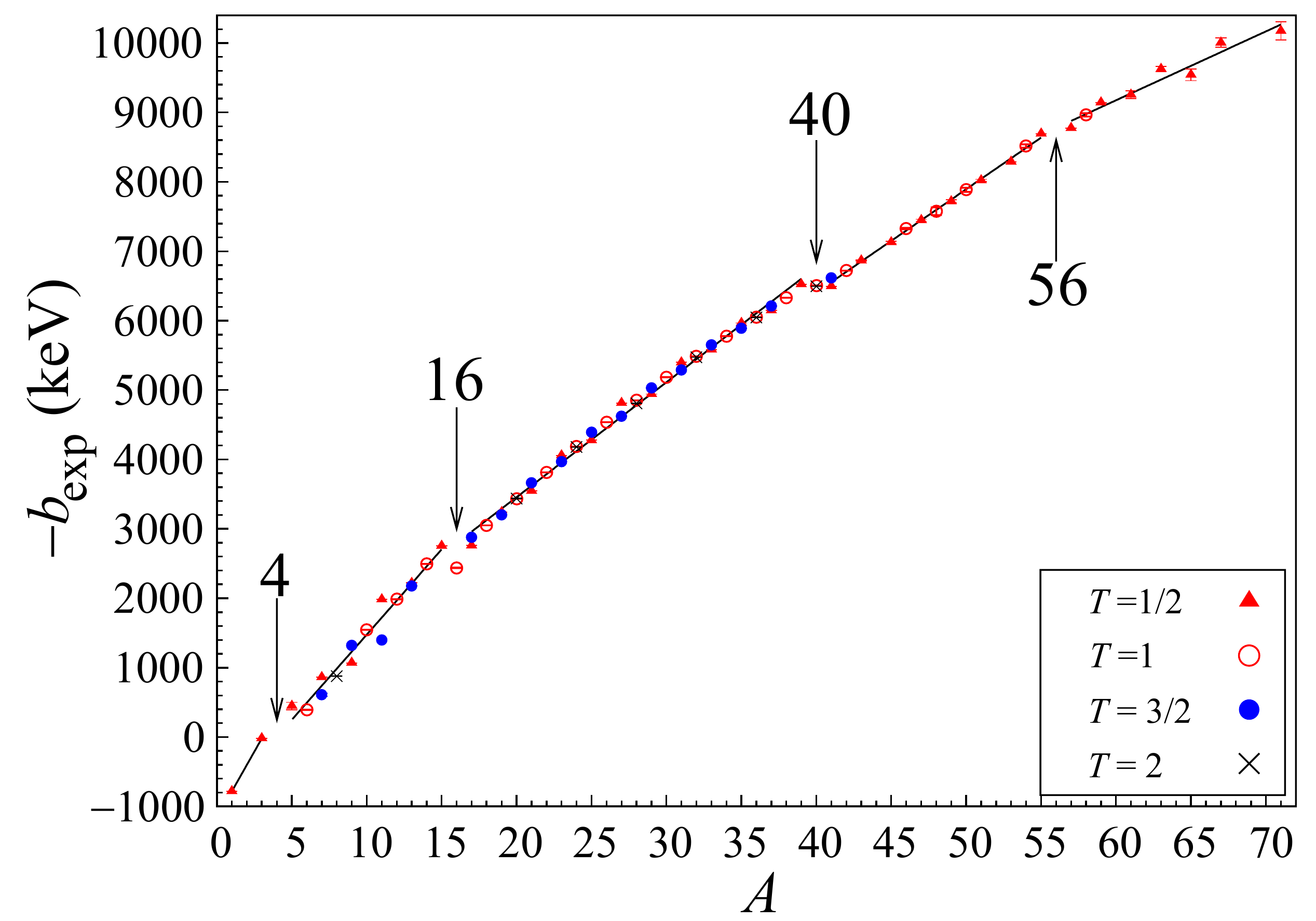}} 
\caption{\label{fig:IMME_All_ADNDT_b_gs_2}
\footnotesize
The $b$ coefficients of the quadratic IMME as a function of $A$ for all lowest-lying $T=1/2,1,3/2,2$ multiplets.
}
\vspace{1cm}
\end{figure*}

\begin{figure*}[h!]
\centering
\rotatebox[]{0}{\includegraphics[scale=0.5, angle=0]{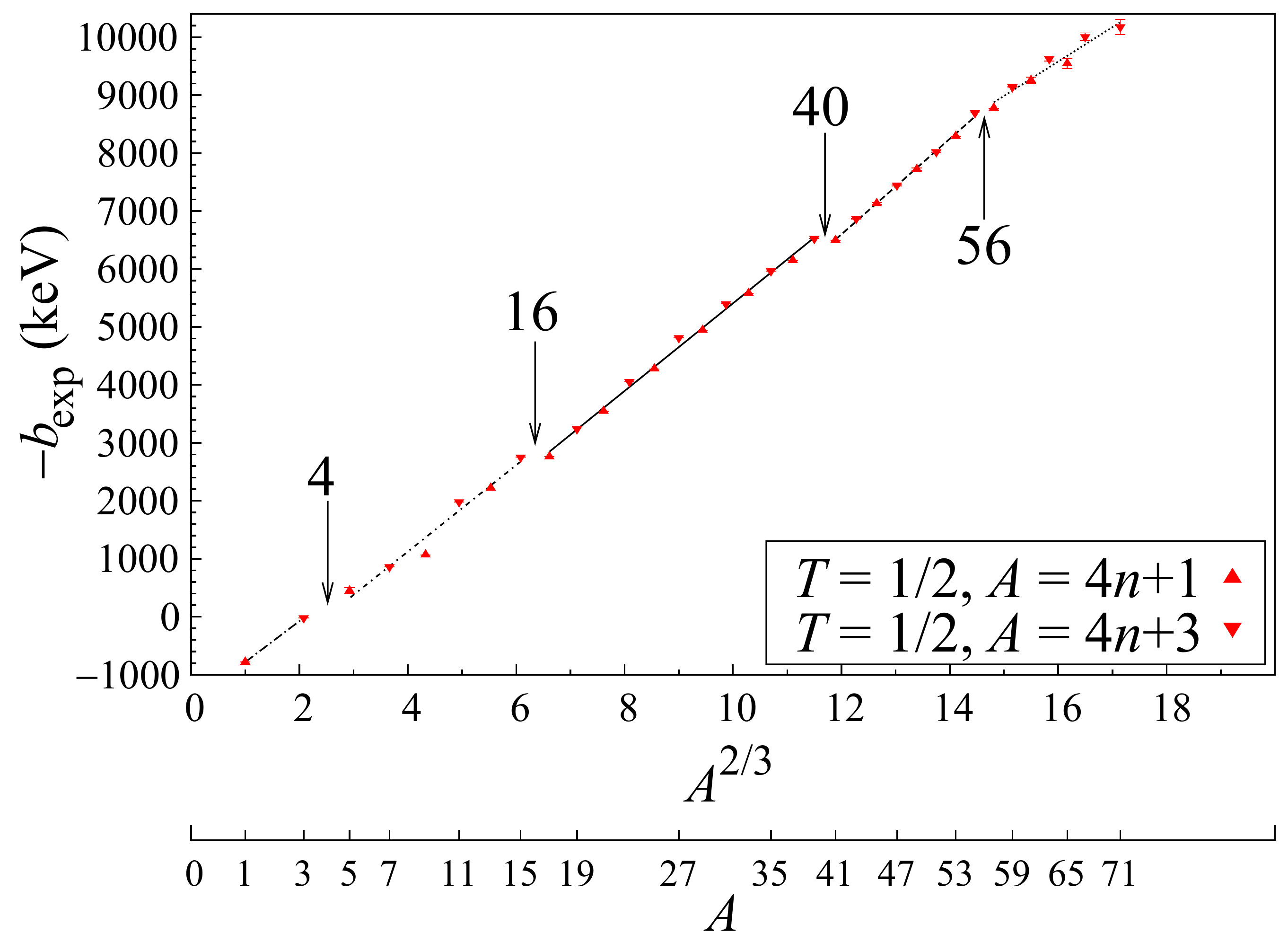}} 
\caption{\label{fig:IMME_ADNDT_b_gs_Thalf}
\footnotesize
The $b$ coefficients of the quadratic IMME as a function of $A^{2/3}$ for all ground states of doublets. 
Unweighted fits within various shell spaces are shown by lines: 
Dotted-dashed line, $s$ shell space, $b=-707.12A^{2/3} + 1489.5$ (keV), $A=1,3$. 
Dotted-short-dashed line, $p$ shell space, $b= -746.51A^{2/3} + 1861.7$ (keV), $A=5,7,9,11,13,15$. 
Solid lines, $sd$ shell space, $b=-756.1A^{2/3} + 2152.3$ (keV), $A=17,19,\ldots,39$. 
Dashed line, $f_{7/2}$ shell space, $b=-824.51A^{2/3} + 3296.4$ (keV), $A=41,43,\ldots,55$. 
Dotted line, $pf$ shell space, $b=-592.77A^{2/3} - 94.754$ (keV), $A=57,59,\ldots,71$.
}
\end{figure*}

\begin{figure*}[h!]
\centering
\rotatebox[]{0}{\includegraphics[scale=0.5, angle=0]{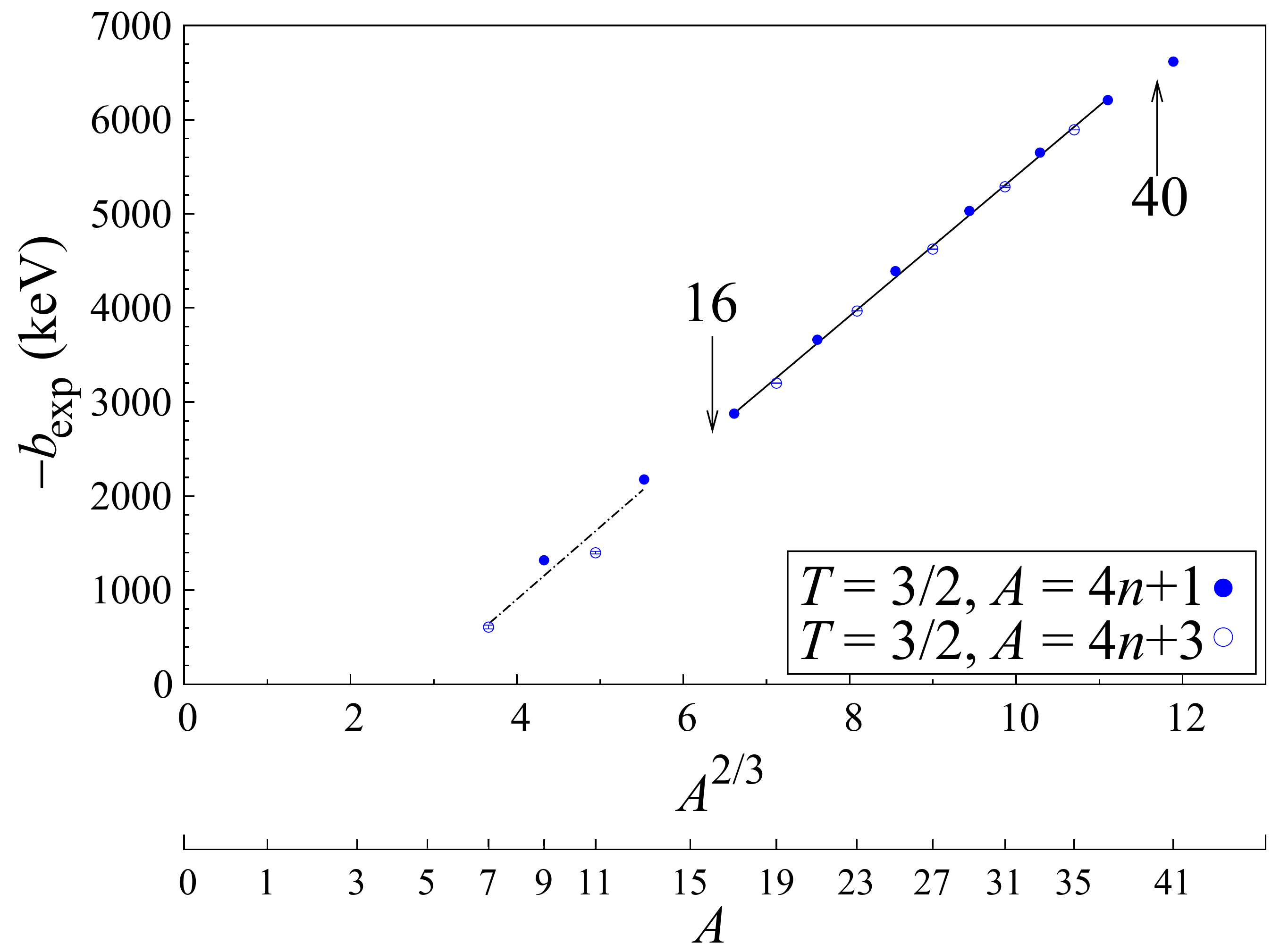}} 
\caption{\label{fig:IMME_ADNDT_b_gs_T3half}
\footnotesize
The $b$ coefficients obtained in a fit to the quadratic IMME for all lowest-lying quartets as a function of $A^{2/3}$ .
Unweighted fits within various shell spaces are depicted by lines: 
Dotted-dashed line, $p$ shell space, $b=-767.76A^{2/3} + 2167.6$ (keV), $A=7,9,11,13$. 
Solid lines, $sd$ shell space, $b=-744.76A^{2/3} + 2044.4$ (keV), $A=17,19,\ldots,37$.
}
\end{figure*}

\begin{figure*}[ht!]
\centering
\rotatebox[]{0}{\includegraphics[scale=0.5, angle=0]{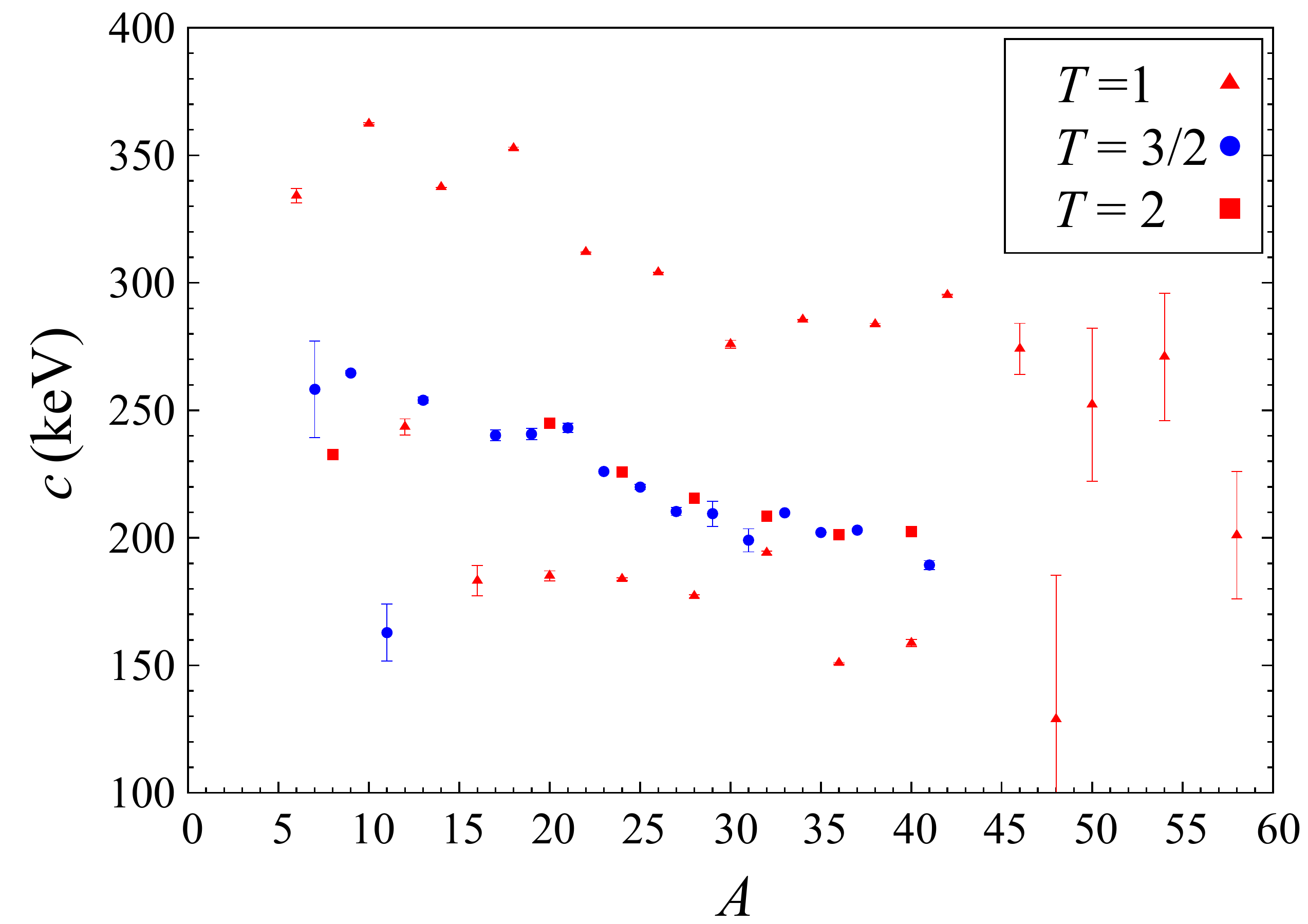}} 
\caption{\label{fig:All_c_coeff}
\footnotesize
The $c$ coefficients of the quadratic IMME as a function of $A$ for all lowest-lying $T=1,3/2$, and 2 multiplets.
}
\vspace{1cm}
\end{figure*}

Over recent years, more experimental data of higher precision on nuclear mass excess and level schemes have been accumulated for most of the $N \approx Z$ nuclei, in particular, for nuclei with mass number ranging from $A=41$ to 71. Incorporating all recent mass measurements from the evaluation~\cite{AME11a} and experimental level schemes~\cite{NNDConline}, we have revised and extended the database of IMME coefficients compiled previously by Britz {\it et al.}~\cite{Britz98}. The total number of multiplets presented in our work incorporates 382 doublets, 132 triplets, 25 quartets and 7 quintets. In particular, it includes recent experimental data on $pf$-shell nuclei. This new set of IMME $a$, $b$, $c$ (and $d$, $e$, defined below in Section ~\ref{sec:Extended_IMME}) coefficients is listed in Tables ~\ref{tab:Doublets} -- ~\ref{tab:Quintets}\footnote{Electronic tables of the IMME coefficients are available at \url{https://github.com/lamyihua/isobaric-multiplet-mass-equation}}. The $b$, $c$ and $d$ coefficients (sometimes for the lowest-lying multiplets only) are also shown in Figs.~\ref{fig:All_b_coeff} -- \ref{fig:IMME_ADNDT_b_gs_T3half}, Figs.~\ref{fig:All_c_coeff} -- \ref{fig:IMME_c_Staggering} and Fig.~\ref{fig:All_d_coeff}, respectively. 

As seen from Eq.~(\ref{eq:IMME}), $a$ and $b$ coefficients for doublets\footnote{There are no $c$ coefficients for doublets.} and $a$, $b$, and $c$ coefficients for triplets can be determined in a unique way from the exact solution of a system of two and three linear equations for two or three multiplet members, respectively. For triplets, the $a$ coefficient is equal to the mass excess of the $T_z=0$ nucleus, so it is not mentioned separately in Table 2.

In principle, knowledge of three mass excesses from a given quartet or quintet is also sufficient to determine the $a$, $b$ and $c$ coefficients of the quadratic IMME. However, we do not consider such incomplete multiplets. In the present compilation an isobaric multiplet is taken into consideration only if the mass excesses of all multiplet members are known experimentally. To this end, the $a$, $b$ and $c$ coefficients of the quadratic IMME, Eq.~(\ref{eq:IMME}), are obtained by a least-square fit to four or five mass excesses for a quartet or for a quintet, respectively.

\bigskip
\newpage
\section{IMME $b$ and $c$ Coefficients}
\label{sec:IMME_bNc}

\subsection{Uniformly charged sphere estimates}

Before we discuss the trends of the experimental $b$ and $c$ coefficients, let us remind a simple model prediction. If we assume that the Coulomb interaction is the only contribution shifting isobaric-analogue states (IAS), forming an isobaric multiplet, and treat a nucleus as a uniformly charged sphere of radius $R=r_0 A^{1/3}$, the total Coulomb energy of a nucleus is given by
\small
\begin{equation}
\label{eq:chargedSphere}
	E_{coul} = \frac{3e^2}{5 R} Z(Z-1) = \frac{3e^2}{5 r_0 A^{\frac{1}{3}}} \left[ \frac{A}{4}(A-2) + (1-A)T_z + T_z^2 \right] \, .
\end{equation}
\normalsize
Putting this expression into the IMME form, one can get the following estimates of the IMME $b$ and $c$ coefficients~\cite{BetheBacher1936,Benenson1979,BentleyLenzi2007}:
\small
\begin{equation}
\label{eq:chargedSphere_abc}
	b = -\frac{3e^2}{5 r_0 } \frac{(A-1)}{A^{\frac{1}{3}}} \, , \,\,\,\,\, 
	c = \frac{3e^2}{5 r_0 } \frac{1}{A^{\frac{1}{3}}} \, , \,\,\,\,\, 
\end{equation}
\normalsize
where $e^2 = 1.44$ MeV$\cdot$fm.

Using the approximation $Z(Z-1) \approx Z^2$, one can get instead of the first equation in Eq.~(\ref{eq:chargedSphere_abc}) an even simpler, often used expression for the $b$-coefficient~\cite{Janecke1966a}:
\small
\begin{equation}
\label{eq:chargedSphere_b}
	b=-\frac{3 e^2}{5 r_0}A^{2/3} \, .
\end{equation}
\normalsize

\begin{figure*}[ht!]
\centering
\rotatebox[]{0}{\includegraphics[scale=0.5, angle=0]{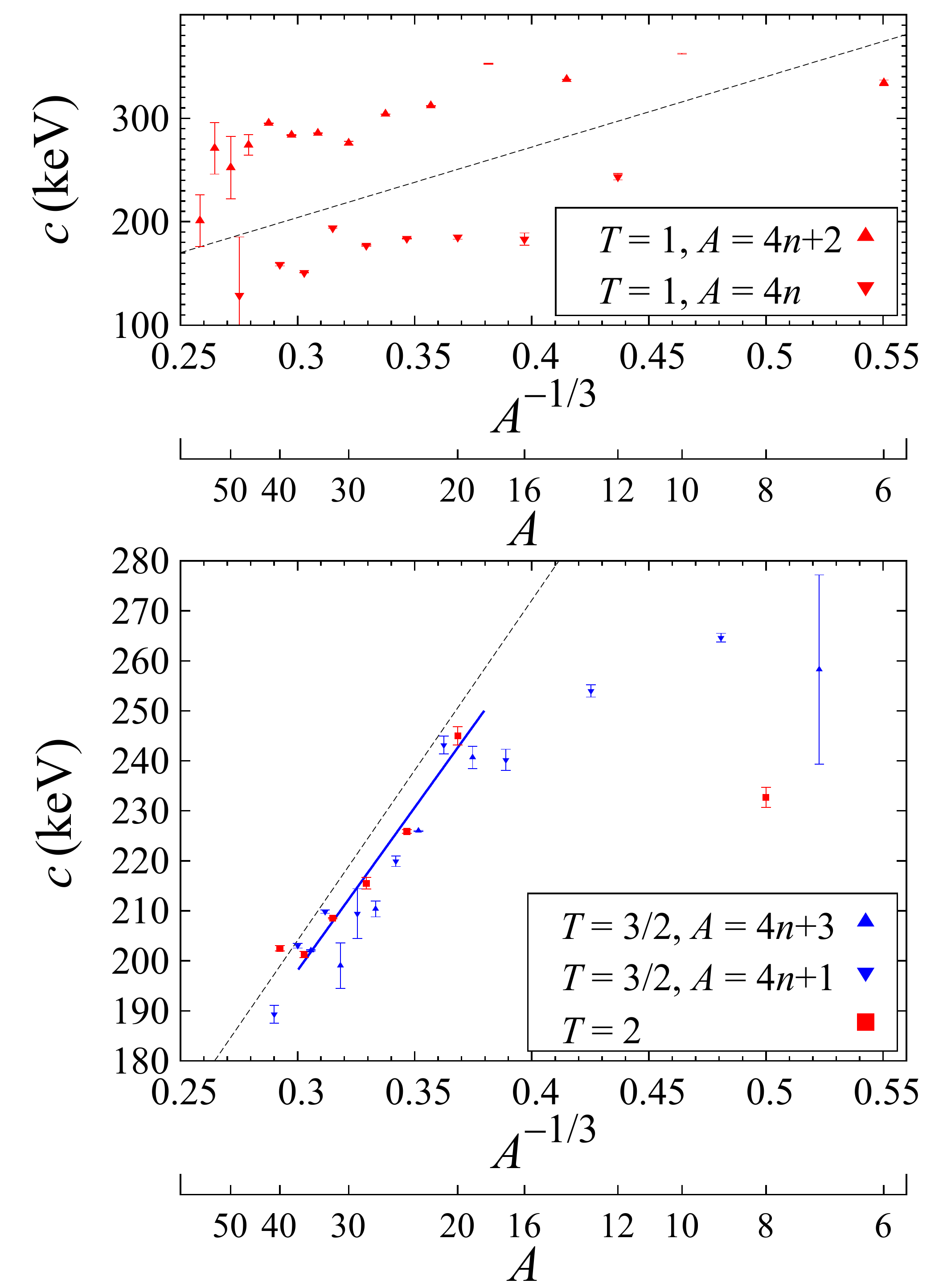}} 
\caption{\label{fig:IMME_All_c_gs_02}
\footnotesize
The $c$ coefficients of the quadratic IMME as a function of $A^{-1/3}$ for the lowest-lying triplets, quartets, and quintets. 
The solid (blue) line is an unweighted fit to the quintets of $A=20, 24, 28, 32$, and 36, $c=643.9A^{-{1/3}} + 5.097$ (keV). The dashed lines in the upper and the lower figures are $c = \frac{3e^2}{5 r_0 } A^{-\frac{1}{3}}$.
}
\end{figure*}

\begin{figure*}[h!]
\centering
\rotatebox[]{0}{\includegraphics[scale=0.5,angle=0]{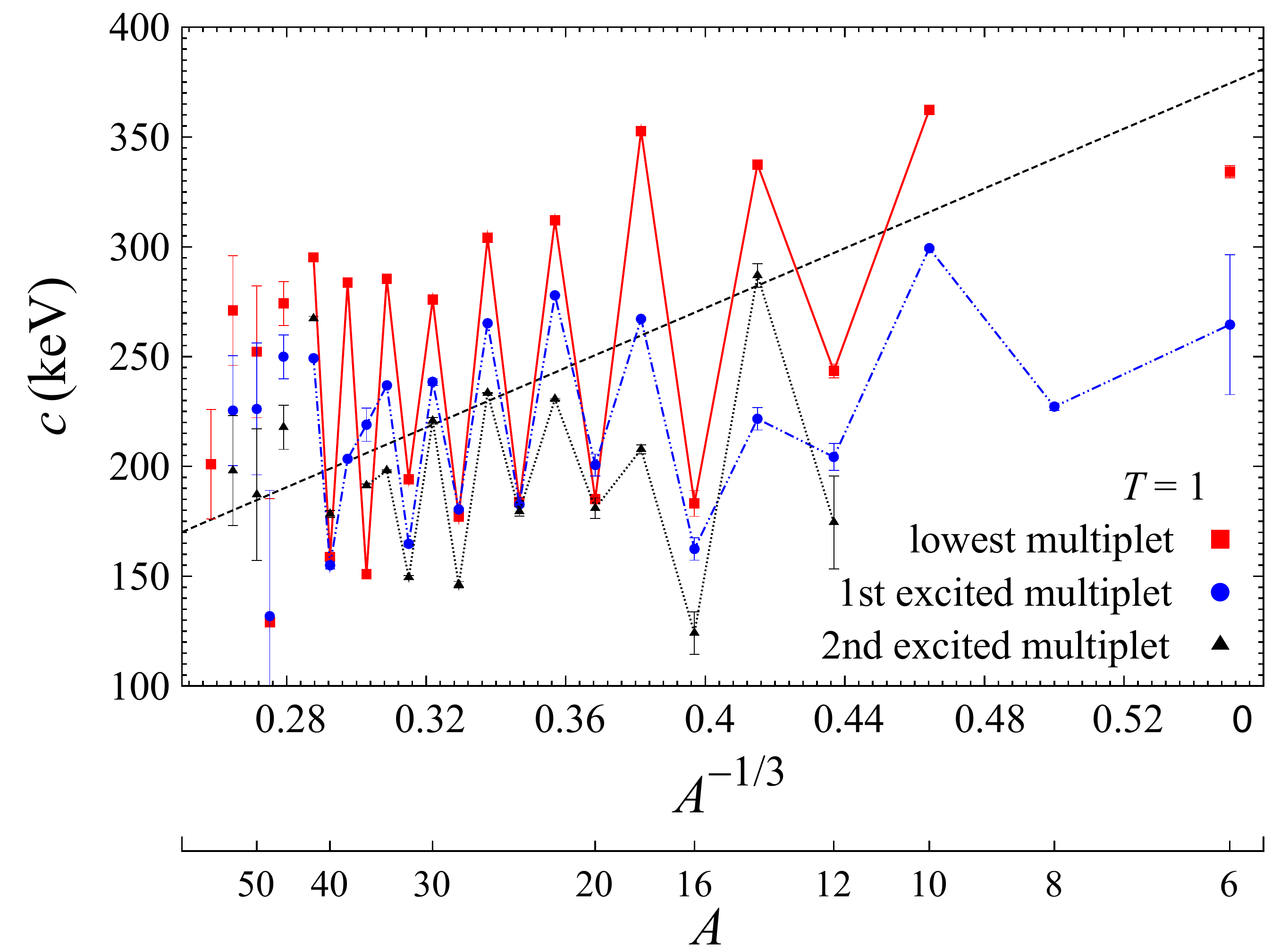}} 
\caption{\label{fig:IMME_c_Staggering}
\footnotesize
The $c$ coefficients of the lowest, first excited and second excited triplets as a function of $A^{-1/3}$. 
The solid (red) line connects the lowest-lying triplets' $c$ coefficients. The (blue) double-dotted-dashed line links (blue) dots which display the first higher-lying triplets' $c$ coefficients. The dotted line links triangles which represent the second higher-lying triplets' $c$ coefficients. The (black) dashed line is $c=\frac{3e^2}{5 r_0 }A^{-1/3}$.
}
\end{figure*}

These crude estimates of the Coulomb contribution to the $b$ and $c$ coefficients are shown in Fig.~\ref{fig:All_b_coeff} and Fig.~\ref{fig:IMME_All_c_gs_02} -- \ref{fig:IMME_c_Staggering}, respectively.

\subsection{IMME $b$ Coefficients}
\label{sec:IMME_b}
		
Fig.~\ref{fig:All_b_coeff} shows the values of $b$ coefficients with an opposite sign ($-b$ values) as a function of $A^{2/3}$. These $b$ coefficients are defined in Eq.~(\ref{eq:IMME}) and thus they are determined exactly for doublets and triplets, while they are established by a least-square fit for quartets and quintets. The numerical values are summarised in Tables ~\ref{tab:Doublets} -- \ref{tab:Triplets} for doublets and triplets, respectively. They are given in the 7th column of Table~\ref{tab:Quartets} and in the 6th columns of Table~\ref{tab:Quintets} for quartets and quintets, respectively. As seen from the plot, the estimates of $b$ coefficients from a uniformly charged sphere, Eq.~(\ref{eq:chargedSphere_abc}) and Eq.~(\ref{eq:chargedSphere_b}), predict\ correctly the overall trend of the $b$ experimental coefficients, underestimating their magnitude by about \mytilde 1 MeV. Bethe and Bacher \cite{BetheBacher1936} proposed that the effect of antisymmetrisation should be considered instead of a purely classical estimate of the Coulomb energy in a nucleus. Based on this idea, Sengupta calculated the Coulomb energy of mirror nuclei with $A=3,\cdots, 39$ using a statistical model \cite{Sengupta1960}. In their extensive review on Coulomb energies~\cite{Nolen_Schiffer69}, Nolen and Schiffer analysed various theoretical approaches, including results of Ref.~\cite{Sengupta1960}, and showed that calculations of Coulomb energies of nuclei in $p$, $sd$ and $pf$-shell spaces underestimate experimental data systematically by about 5\% -- 10\%.	They concluded that the Coulomb interaction alone is not sufficient to reproduce the experimental isobaric multiplet splittings and therefore non-Coulomb charge-dependent forces might be considered (an empirical approach within the shell model can be found in Refs.~\cite{OrBr89,YiHuaNadya2012a, YiHuaThesis}).

The solid line in Fig.~\ref{fig:All_b_coeff} corresponds to the weighted fit of all $b$ coefficients, $b=-691.39(90) A^{2/3} + 1473.44(95)$ (keV). Most of the lowest-lying multiplets have a lower error bar than the higher-lying multiplets. Hence, the weighted fit line is closer to the $b$ coefficients of the lowest lying multiplets. However, the unweighted fit (dashed line) function, $b=-723.4 A^{2/3} + 1927.9$ (keV), presents a better overall average description of $b$ coefficients. The discrepancy between the weighted fit and the unweighted one shows that the absolute values of $b$ coefficients tend to decrease with excitation energy for higher lying multiplets. The unweighted fit is less influenced by the few low-$A$ nuclei which have more precise measurements on nuclear mass and energy level schemes. A method of systematically analysing the behaviour of the excitation energy differences in doublets was given by Everling ~\cite{Everling1970}.

Fig.~\ref{fig:IMME_All_ADNDT_b_gs_2} presents $b$ coefficients of  all known lowest-lying multiplets as a function of $A$. We found out that the dependence can be extrapolated piece-wise by straight lines. The best fits are given in the figure. Discontinuities of these fitted lines occur at the closed $(p,sd,pf)$ shells at $A=$4, 16, and 40, which was also noticed earlier by J\"{a}necke ~\cite{Janecke1966a}. These shell effects are not described by the uniformly charged sphere assumption. In the present study, we find that there is also a discontinuity at $A=56$; it points toward a $0f_{7/2}$ subshell gap. Furthermore, we find that discontinuities are also inherent for quartets, which was not evidenced in Ref.~\cite{Janecke1966a} due to limited experimental data.

Figs.~\ref{fig:IMME_ADNDT_b_gs_Thalf} -- \ref{fig:IMME_ADNDT_b_gs_T3half} depict separately the $b$ coefficients of the lowest-lying doublets and quartets, respectively, as a function of $A^{2/3}$. The data on doublets spread up to $A=71$, while the data for quartets are known only up to $A=41$. Straight lines represent the best fits to data. In these plots, it can be well noticed that besides the above mentioned discontinuities, the data exhibit a well pronounced staggering effect. The $b$ coefficients form clearly two families of multiplets for $A=4n+1$ and $A=4n+3$, which lie slightly below or above the fitted lines, respectively. No staggering in the dependence of the $b$ coefficients for triplets or quintets can be noticed. These small-amplitude oscillations of $b$ coefficients can be related to the Coulomb contribution to the pairing energy (see Refs.~\cite{Janecke1966a,YiHuaNadya2012a} for detail).

\subsection{IMME $c$ Coefficients}
\label{sec:IMME_c}

The experimental $c$ coefficients of the quadratic IMME of all known lowest-lying $T=1,3/2$ and 2 multiplets are plotted in Fig.~\ref{fig:All_c_coeff} as a function of $A$. For triplets, the $c$ coefficients are calculated exactly from three nuclear mass excesses, while for quartets and quintets, these $c$ coefficients are obtained from a least-square fit to the corresponding four or five members mass excesses. The numerical values are tabulated in Table~\ref{tab:Triplets} for triplets and they can be found in the 7th column of Table~\ref{tab:Quartets} and in the 6th columns of Table~\ref{tab:Quintets} for quartets and quintets, respectively. Fig.~\ref{fig:IMME_All_c_gs_02}  shows separately the experimental $c$ coefficients for the lowest-lying triplets (upper panel) and quartets and quintets (lower panel)  as a function of $A^{-{1/3}}$. Dashed straight lines indicated in the figures represent the estimate deduced from a classical homogeneously charged sphere, Eq.~(\ref{eq:chargedSphere_abc}). For the lowest-lying triplets' and quintets' $c$ coefficients, indicated as red triangles and red squares, respectively, this classical assumption is roughly valid. However, no clear dependence can be seen for quartets (Fig.~\ref{fig:IMME_All_c_gs_02}, lower panel), or for the known higher-lying quartets' $c$ coefficients (not shown in the figures).
	
Figs.~\ref{fig:All_c_coeff} -- \ref{fig:IMME_All_c_gs_02} evidence that the $c$ coefficients of triplets having $A=4n$ and $A=4n+2$ form two distinct families. A regular staggering effect is clearly visible. Quintets in the $sd$-shell space also show a small staggering behaviour when they are plotted as a function of $A^{-{1/3}}$ in Fig.~\ref{fig:IMME_All_c_gs_02}. No oscillatory behaviour can be noticed in quartets' $c$ coefficients.

The staggering also takes place in higher lying triplets, as shown in Fig.~\ref{fig:IMME_c_Staggering}. In this figure, we connect by solid, dashed-dotted and dotted lines the lowest, first and second excited triplets, respectively. These lines stop as soon as there are breaks in experimental data points (e.g., absence of data for $A=44, 52, 56$ triplets). In addition, the ground state multiplet ($J^\pi =2^+$) of the $A=8$ triplet is not considered in our work due to well confirmed large isospin mixing between two neighbouring $2^+$ states in $^8$Be at 16.626 MeV and at 16.922 MeV excitation energy. Two higher lying triplets, $J^\pi =1^+$ and $J^\pi =3^+$, are considered, however, their average energy indicated in Table~\ref{tab:Triplets} should be taken with caution due to the undetermined energy of the $J^\pi =2^+$ $(T=1)$ state in $^8$Be (the average energies of $1^+$ and $3^+$ given in the table are obtained using the 16.626 MeV state in $^8$Be).

Details and a theoretical description of these effects related to the Coulomb contribution to the pairing can be found in Ref.~\cite{Janecke1966a,YiHuaNadya2012c, YiHuaThesis}.

\begin{figure*}[ht!]
\centering
\rotatebox[]{0}{\includegraphics[scale=0.5, angle=0]{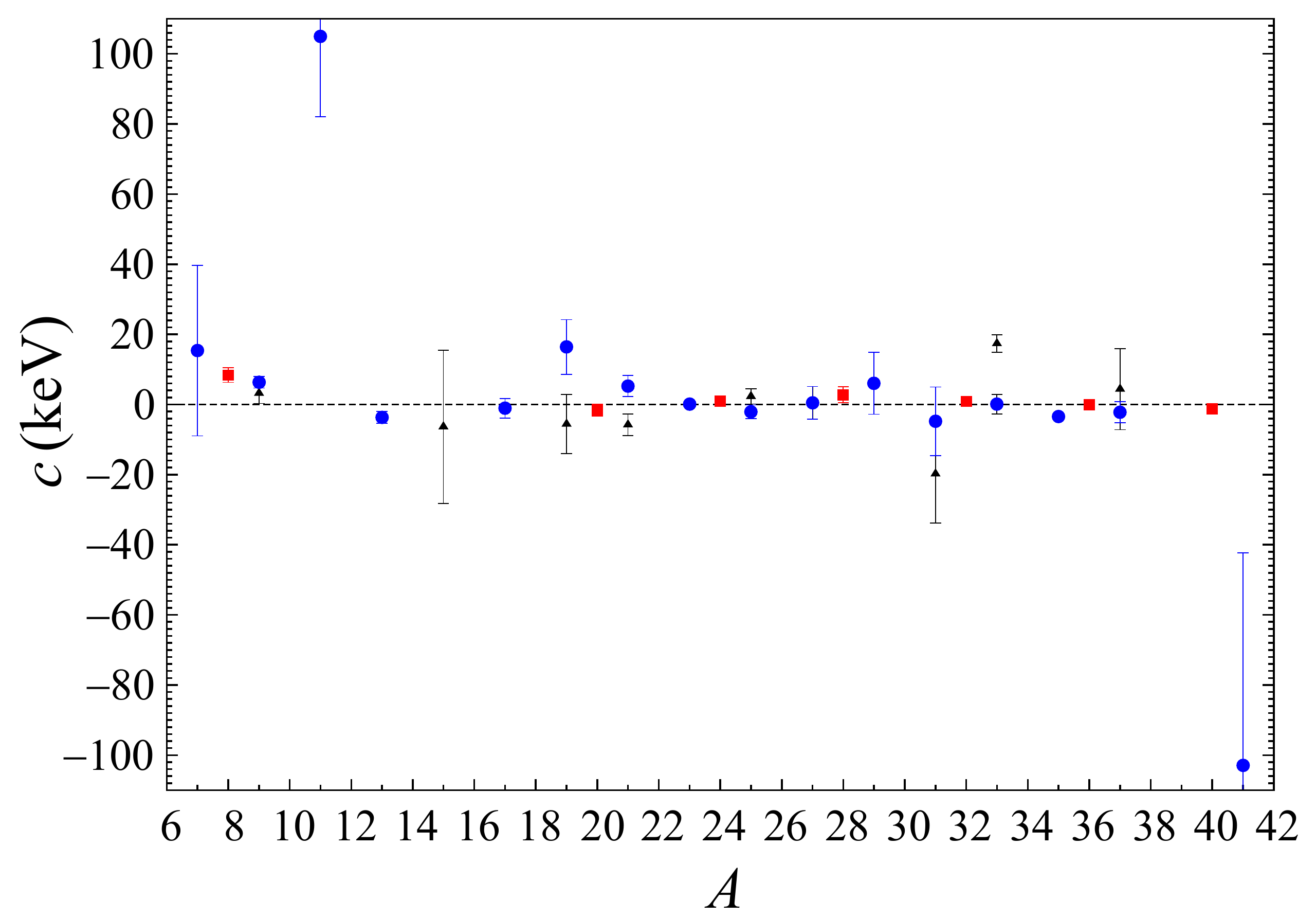}} 
\caption{\label{fig:All_d_coeff}
\footnotesize
The experimental $d$ coefficients as a function of $A$ for all quartets and quintets. 
These $d$ coefficients are defined by the cubic IMME (Eq.~(\ref{eq:IMME_ext}) with $e=0$). The (blue) dots and (red) squares are $d$ coefficients of the lowest-lying quartets and quintets, respectively; whereas (black) triangles are $d$ coefficients of higher-lying quartets. 
}
\end{figure*}	
    
\section{Extended IMME}
\label{sec:Extended_IMME}

The analysis of quartet and quintet mass excesses can be performed within the extended IMME, which includes higher order degrees of $T_z$:
\small
\begin{align}
\label{eq:IMME_ext}
	M(\alpha,T,T_z) &= a(\alpha, T) + b(\alpha, T) T_z + c(\alpha, T) T_z^2 \notag \\
    &+ d(\alpha, T) T_z^3+ e(\alpha, T) T_z^4 \, .
\end{align}	
\normalsize
The last two terms in the right-hand side of this equation may appear due to the higher-body components of the charge-dependent interaction and/or due to the isospin mixing with neighbouring states. In particular, taking into account higher-order perturbation effects ~\cite{Henley_Lacy69} may result in an around 1 keV estimate for a non-zero $d$ coefficient.

Let us notice that a cubic term can be uniquely determined for quartets as
\small
\begin{align}
\label{eq:IMME_d}
	d &= \frac16 \{ M\left(\alpha, T, T_z=3/2\right) -M\left(\alpha, T, T_z=-3/2\right)  \notag \\
	&-3\left[M\left(\alpha, T, T_z=1/2\right) -M\left(\alpha, T, T_z=-1/2\right)\right] \} \, .
\end{align}	
\normalsize
Similarly, both $d$ and $e$ can be extracted from a five-member quintet as
\small
\begin{align}
\label{eq:IMME_de}
	d &= \displaystyle \frac{1}{12} \{M\left(\alpha, T, T_z=2\right) -M\left(\alpha, T, T_z=-2\right) \notag \\
	&+2\left[M\left(\alpha, T, T_z=1\right) -M\left(\alpha, T, T_z=-1\right)\right] \} \notag \\
	e &= \displaystyle \frac{1}{24} \{ M\left(\alpha, T, T_z=2\right) +M\left(\alpha, T, T_z=-2\right) \notag \\
	&-4\left[M\left(\alpha, T, T_z=1\right) +M\left(\alpha, T, T_z=-1\right)\right] \notag \\
    &+6M\left(\alpha ,T,T_z=0\right) \} 
\end{align}	
\normalsize

Table ~\ref{tab:Quartets} shows the $a$, $b$, and $c$ coefficients deduced from the quadratic IMME fit to $T=3/2$ quartets, together with the normalised value of the $\chi^2$-deviation. The last column presents the exact determination of the $a$, $b$, $c$ and $d$ coefficients from experimental mass excesses.

It can be noticed that most of the $d$ coefficients of quartets 
are larger than 1 keV (from \mytilde2 keV to \mytilde100 keV) and necessitate much lower error bars, c.f. Fig.~\ref{fig:All_d_coeff}.
More precise measurements of the masses and excitation energies are certainly required to constrain the values of $d$ coefficients.

The only clearly non-zero values are confirmed for the $A=9$ and $A=35$ ground state $J^\pi =3/2^-$ quartets. As was shown in a recent work~\cite{Brodeur12}, a large value of the $d$ coefficient in the $A=9$ ground state quartet is most probably related to a strong isospin mixing with a high-lying $T=1/2$ state.

Table~\ref{tab:Quintets} lists the fitted IMME coefficients for quintets. The table includes results of the quadratic IMME fit following Eq.~(\ref{eq:IMME}) and given in the 6th column, as well as the extended IMME fit  following Eq.~(\ref{eq:IMME_ext}) where either a non-zero cubic or a non-zero quartic term is considered (columns 7 and 8, respectively). The last column of Table~\ref{tab:Quintets} contains $a$, $b$, $c$, $d$ and $e$ coefficients of the full five-term IMME, Eq.~(\ref{eq:IMME_ext}), obtained from the exact solution of a system of five equations describing mass excesses of all five members of a given quintet.
 
The $d$ values corresponding to the cubic fit for quintets are shown in Fig.~\ref{fig:All_d_coeff}, together with $d$ coefficients for quartets. From the normalised $\chi^2$ value of the quadratic IMME fit for $A=8$ (quintet), there is a strong indication for the need of the cubic term. This quintet has the smallest $A$ of all known quintets, but it has the highest $d$ value. It may be due to non-perturbative effects which are caused by the less tightly bound nature and small Coulomb barriers of those states ~\cite{Triambak2006}.	

The non-zero $d$ coefficient for the lowest-lying $A=32$ quintet is confirmed ($d=0.83 \pm0.22$ keV). From a shell model study, it arises most probably from the isospin-mixing with nearby states~\cite{SignBrown11}.
	
\section{Summary}
\label{sec:summary}

We present a new compilation of experimental IMME coefficients based on the recently compiled mass excesses ~\cite{AME11a} and energy levels from the evaluated nuclear structure data files of the National Nuclear Data Center (up to end of 2011). The database comprises 546 multiplets. The behaviour of $b$ and $c$ coefficients as a function of nuclear mass number is studied in detail. A pronounced staggering in the values of $b$ coefficients is reported for doublets and quartets, while no similar effect is noticed for the triplets' $b$ coefficients. At the same time, an analysis of $c$ coefficients evidences the existence of staggering for triplets (for the lowest-lying and even first and second excited multiplets), while any oscillatory behaviour is absent in the trend of quartets' $c$ coefficients. These effects are related to the Coulomb contribution to the pairing and are studied elsewhere.

Exploration for the IMME beyond the quadratic form confirms non-zero values of $d$ coefficients for $A=9, 35$ quartets and the $A=32$ quintet. As shown by a shell-model analysis, these values are most probably related to the isospin mixing with nearby states. More precise data on quartets and quintets are required to verify and to generalise this hypothesis.

\section*{Acknowledgements}

We are grateful to M. Wang, S. Triambak, and J. Giovinazzo for suggestions on error estimates, and P. Van Isacker for his interest and stimulating discussions. Y. H. Lam gratefully acknowledges partial financial support from the French Embassy in Malaysia (dossiers n$^\circ$ 657426B and n$^\circ$ 703786D) which allowed for a sustainable stay in Bordeaux during the work on this paper. J. B. Bueb and M. S. Antony are grateful to Christelle Roy (Directress of IPHC, Strasbourg) for her encouragement.

\newpage
\onecolumn
\section*{\label{sec:Notations} Notations for Tables~\ref{tab:Doublets}, ~\ref{tab:Triplets}, ~\ref{tab:Quartets}, and ~\ref{tab:Quintets} }


\end{center}

\normalsize

\end{document}